\documentclass[12pt]{iopart}

\usepackage{graphicx}
\newcommand{\be}{\begin{equation}} 
\newcommand{\ee}{\end{equation}} 
\newcommand{\bea}{\begin{eqnarray}} 
\newcommand{\eea}{\end{eqnarray}} 
\newcommand{\eml}{\end{mathletters}} 
\newcommand{\nn}{\nonumber\\} 
\newcommand{\oh}{\frac{1}{2}}

\begin{document} 

\title{Systematics of neutron emission}

\author{D.S. Delion $^{1,2}$ and S. Ghinescu $^{1,3}$}
\address{$^1$ "Horia Hulubei" National Institute of Physics and Nuclear Engineering, 30 Reactorului, POB MG-6, RO-077125, Bucharest-M\u agurele, Rom\^ania}
\address{$^2$ Academy of Romanian Scientists, 3 Ilfov RO-050044, Bucharest, Rom\^ania}
\address{$^3$ Department of Physics, University of Bucharest, 405 Atomistilor, POB MG-11, RO-077125, Bucharest-M\u agurele, Rom\^{a}nia}
\date{\today}

\begin{abstract}
Neutron physics is one of the oldest branches of the experimental nuclear physics,
but the investigation of the spontaneous neutron emission from the ground state along the neutron dripline
is still at its beginning, in spite of the crucial importance for nuclear astrophysics.
The proton dripline is much better investigated and a systematics of spontaneous proton half lives
corected by the centrifugal barrier (monopole transitions) is given by the Geiger-Nuttall law $\log_{10}T\sim\chi$, 
where $\chi\sim ZQ^{-1/2}$ is the Coulomb parameter characterizing the outgoing Coulomb-Hankel wave 
in terms of the daughter charge $Z$ and Q-value.
Our purpose is to propose a similar simple systematics of spontaneous neutron half lives, but in terms 
of the nuclear reduced radius $\rho=\kappa R\sim A^{1/3}Q^{1/2}$, 
characterizing the "neutral" outgoing spherical Hankel wave.
It turns out that the half life in emission of neutral particles is governed by the scaling law 
$T\sim\rho^{-2}\sim A^{-2/3}Q^{-1}$ for monopole transitions.
We evidence the important role of the angular momentum carried by the emitted neutron.
The influence of the neutron wave function generated by a Woods-Saxon nuclear mean field is also analyzed.
\end{abstract}


\maketitle

\section{Introduction} 
\label{sec:intro} 
\setcounter{equation}{0}
\renewcommand{\theequation}{1.\arabic{equation}}

The theoretical description of charged particle emission from nuclei, like protons, $\alpha$-particles, 
or heavy clusters, is described in terms of the quantum penetration through the Coulomb barrier 
surrounding the nuclear interaction between the preformed cluster and remaining (daughter) nucleus \cite{Del10}.
It is based on the work published by Gamow \cite{Gam28} and independently by Condon and Gurney \cite{Con28}, 
almost a century ago. The main conclusion extrated from the analysis of emission data can be summarized
in a Geiger-Nuttall type of law connecting the logarithm of the half life, or decay width, to the Coulomb parameter.
The so-called reduced width is obtained by dividing the decay width to the Coulomb barrier penetration \cite{Del10}.
The other important law of emission processes linearly connects the logarithm of this quantity to the fragmentation 
potential, defined by the difference between the Coulomb barrier and Q-value \cite{Del09,Ghi21,Dum22}.

The limits of nuclear stability versus particle emission known as proton or neutron driplines, where the last
proton/neutron becomes unstable, were intensively investigated from both
experimental \cite{Han03} and theoretical sides \cite{Erl12,Neu20}.
Proton and two-proton emission processes from the ground state take place in the proton dripline region.
This area is experimentally well known up to Pb region, due to the relatively large half lives 
of protons in continuum induced by the Coulomb barrier \cite{Del06,Del06a}.

On the other hand, modern experimental facilities were able to produce isotopes along the neutron dripline
for elements up to argon \cite{Cot12,Now21}.
In the absence of the Coulomb barrier, the corresponding centrifugal barrier has a weaker bounding
effect on neutrons in continuum and therefore the corresponding half lives are much shorter. 
Neutron separation energy, required to remove to infinity a neutron from a nucleus (Z, N), 
leaving a residual nucleus (Z, N-1) was systematized by several papers \cite{Joh57,Gel60,Ban68}.
It was shown that single- and two-nucleon separation energies can be parametrized in terms of
N/Z ratio and the mass number A \cite{Vog01}. 
This quantity has negative values for neutrons lying in continuum and defines the neutron dripline.
Many theoretical investigations were performed in order to predict the position of the neutron dripline.
The pair correlation energy along the neutron dripline was investigated \cite{Ber91,Hag05,Ham06,Pas13}.
Ground-state properties of nuclei with extreme neutron-to-proton ratios were described in the
framework of the self-consistent mean-field theory with pairing correlations \cite{Dob96,Sto03,Sto13}.
Clustering and molecular states in neutron rich isotopes of Lithium, Beryllium, Boron and Carbon, 
were recently investigated \cite{Gno22}.
The prediction of the two-neutron dripline was analyzed in the framework of covariant density functional 
theory \cite{Afa15}.
Let us mention that the r-process (rapid neutron-capture process) of stellar nucleosynthesis explains the production 
of stable (and some long-lived radioactive) neutron-rich nuclides heavier than iron observed in stars \cite{Arn07}.

In Ref. \cite{Del06} a systematics of spontaneous proton emission half lives was proposed
in terms of the Coulomb parameter ($\chi$), defining the outgoing Coulomb-Hankel wave
\bea
H^{(+)}_l(\chi,\rho)=G_l(\chi,\rho)+iF_l(\chi,\rho)~,
\eea
depending on irregular $G_l(\chi,\rho)$ and regular $F_l(\chi,\rho)$ Coulomb functions with a given angular momentum $l$.
It turns out that the modulus of this function squared is proportional to the proton half life \cite{Del10}.
For neutron emission the Coulomb parameter vanishes, but the other parameter, defining the outgoing
"neutral" spherical Hankel function, is given by the reduced radius $\rho$.
The aim of this letter is to propose a similar systematics of spontaneous neutron half lives, 
but in terms of this quantity.

\section{Theoretical background} 
\label{sec:theor} 
\setcounter{equation}{0} 
\renewcommand{\theequation}{2.\arabic{equation}} 

The theory of neutron emission is an old issue, but we will remind its main details in order
to fix the main concepts.
The standard way to analyze a single particle (proton or neutron) unbound state is to consider it as
a resonant state in continuum. This is an eigenstate of some mean field potential with complex energy 
and outgoing asymptotics described by the Coulomb-Hankel function \cite{Del10}.
For outgoing narrow resonances, called Gamow states,
the real part of the energy is positive and it is called Q-value.
The decay width $\Gamma$ is twice the complex part of the energy and for narrow resonances it satisfies
the condition $\Gamma<<Q$. Thus, the wave function describing the emission process is given by the following ansatz
\bea
\Psi({\bf R},t)=\Phi({\bf R})\exp\left[-\frac{i}{\hbar}\left(Q-i\frac{\Gamma}{2}\right)t\right]~.
\eea
It corresponds to a pole of the S-matrix in the complex energy plane. 

In the alternative description, by using real scattering states with given angular momentum and energy
\bea
&&\psi_l(R,E)\sim G_l(\chi,\rho)\sin\delta_l(E)+F_l(\chi,\rho)\cos_l\delta_l(E)
\nn
&=&\frac{i}{2}e^{-i\delta_l(E)}\left[H^{(-)}_l(\chi,\rho)-S_l(E)H^{(+)}_l(\chi,\rho)\right]~,
\eea
defining the S-matrix $S_l(E)=\exp[2i\delta_l(E)]$, the phase shift for some narrow resonant state 
sharply passes through the value $\delta_l=\pi/2$ where only the irregular component $G_l$ remains.
Thus, for a narrow resonance, where one has $|G_l(\chi,\rho)|>>|F_l(\chi,\rho)|$ inside the barrier \cite{Del10}, 
both complex Gamow and real scattering state descriptions give close results.

The low-lying proton eigenstates in continuum have very small decay widths and
therefore are long living due to the large Coulomb potential.
In this case the equivalent expression of the particle decay width of the process can be expressed 
by using the continuity equation as follows \cite{Del10}
\bea
\label{Gamma}
\Gamma=\hbar v \frac{s_{lj}}{\Omega_j}\vert N_{lj}\vert^2~,
\eea
where $l,j$ denotes the angular momentum and total spin of the emitted particle, $s_{lj}$ the spectroscopic
factor, $\Omega_j=j+\oh$ the level occupancy and $v=\sqrt{2Q/\mu}$ the particle velocity with
the reduced mass $\mu=mm_D/m_P$, written in  terms of the particle, daughter and parent masses.
For superfluid nuclei the spectroscopic factor within the  Bardeen-Cooper-Schieffer (BCS) approximation 
is given by the probability of the particle component in the quasiparticle state at the Fermi level, 
where the single particle energy lies close to the Lagrange multiplier $\epsilon_{lj}\approx\lambda$, 
thus giving $s_{lj}=u^2_{lj}=\oh[1+(\epsilon_{lj}-\lambda)/E_{{lj}}]\approx$ 0.5. 
The quantity $N_{lj}$ is called scattering amplitude and it is given by the ratio 
\bea
\label{N}
N_{lj}=\frac{f_{lj}(R)}{H^{(+)}_l(\chi,\rho)/\rho}~,
\eea
between the internal wave function normalized to unity inside nucleus $f_{lj}(R)$ and external outgoing wave,
given by the Coulomb-Hankel spherical wave $H^{(+)}_l(\chi,\rho)/\rho$, depending upon the Coulomb-Sommerfeld 
parameter $\chi=2Ze^2/(\hbar v)$. Here, Z is the daughter charge, and nuclear reduced radius $\rho=\kappa R$ defined in terms 
of the linear momentum $\kappa=\mu v/\hbar$. 
Let us mention that the above ratio obviously does not depend upon the radius $R$ in the region where
the nuclear interaction vanishes, because both internal and external functions satify the same Schr\"odinger equation.
Let us mention that the same result is obtained within the real scattering approach \cite{Lan58}.

At the same time, some low-lying neutron resonant states have relative large centrifugal 
barriers and therefore are also relative long living. Therefore the decay width can be expressed by (\ref{Gamma}),
but the external outgoing wave for chargeless particles with $\chi=0$ is given by the spherical Hankel function
$H^{(+)}_l(0,\rho)/\rho\equiv h^{(+)}_l(\rho)$ and describes the neutron motion at large distances, 
where the nuclear interaction vanishes.

\begin{figure}[h]
\begin{center} 
\includegraphics[width=10cm]{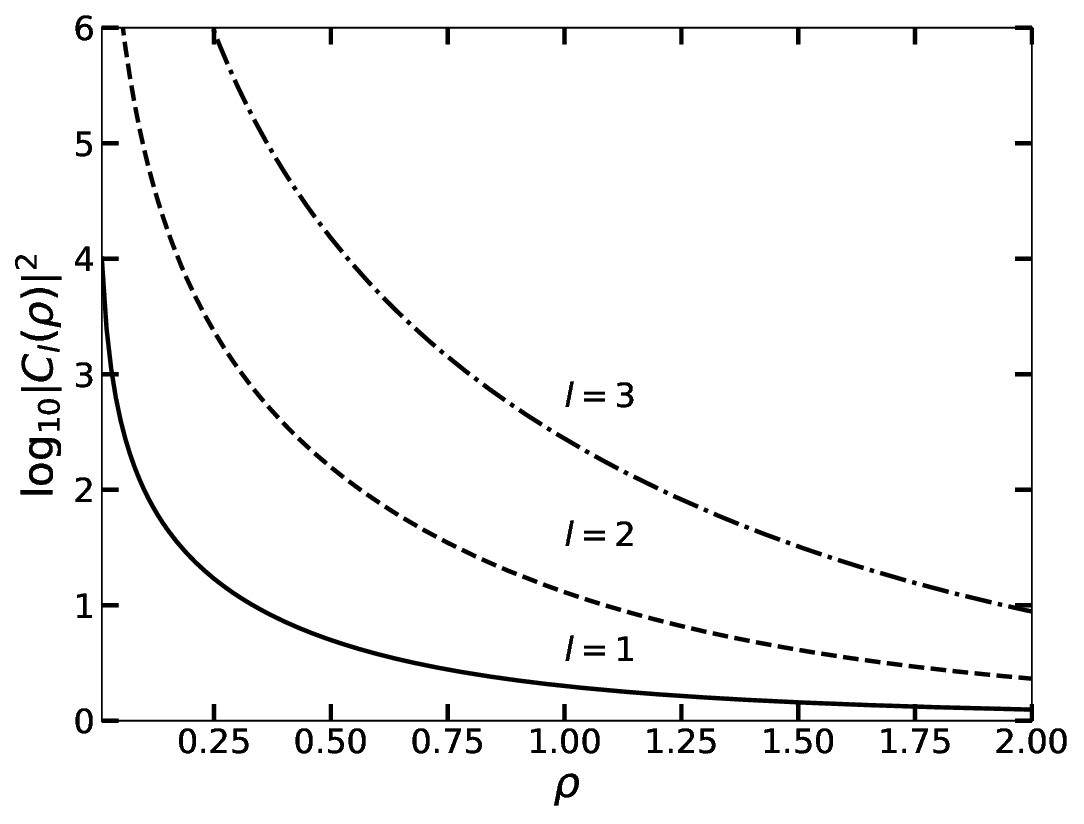} 
\caption{
The values of the centrifugal factor $|C_l(\rho)|^{2}$, given by Eq. (\ref{C}), entering the half life (\ref{T}) 
versus reduced radius $\rho$ for $l$=1 (solid line), 2 (dashed line), 3 (dot-dashed line).
}
\label{fig1}
\end{center} 
\end{figure}

The spherical Hankel function is defined in terms of spherical Neumann and Bessel functions
\bea
h^{(\pm)}_l(\rho)= y_l(\rho)\pm ij_l(\rho)~.
\eea
It turn out that it can be factorized 
\bea
\label{h}
h^{(\pm)}_l(\rho)&=&h^{(\pm)}_0(\rho)C_{l}(\rho)~,
\eea
into monopole ($l=0$) and centrifugal ($l>0$) components
\bea
\label{C}
h^{(\pm)}_0(\rho)&=&\frac{e^{\pm i\rho}}{\rho}
\nn
C_{l}(\rho)&=&\sum_{m=0}^l\frac{i^{m-l}}{m!(2\rho)^m}\frac{(l+m)!}{(l-m)!}~.
\eea

In Fig. \ref{fig1} we plotted the centrifugal factor $|C_{l}(\rho)|^{2}$ 
for $l$=1 (solid line), $l$=2 (dashed line) and $l$=3 (dot-dashed line) versus the reduced radius $\rho$ 
estimated at the geometrical nuclear radius $R_0=1.2A^{1/3}$
\bea
\rho=\kappa R_0=\frac{(2\mu Q)^{1/2}}{\hbar}1.2A^{1/3}\approx 0.263~Q^{1/2}A^{1/3}~,
\eea
written in terms of the neutron Q-value and mass-number $A$. 
From this figure one sees that the centrifugal effect gives an effect up to two orders of magnitude per unit of angular momentum.
Notice that by considering for experimental values $Q<$ 2 MeV, $A<$ 30  one obtains a relative small interval of values
$\rho<1$. 
We also make theoretical predictions by extending the analyzed interval up to $\rho$=2, corresponding to $A=200$,
allowing in this way to predict the half lives of the neutron dripline.

Thus, the theoretical half life can be written as follows
\bea
\label{T}
T=\frac{\hbar\ln 2}{\Gamma}
=\frac{\Omega_j\ln 2}{s_{lj}v}\left|\frac{h^{(+)}_l(R)}{f_{lj}(\rho)}\right|^2
=\frac{\Omega_j\ln 2}{s_{lj}v\rho^2}\left|\frac{C_l(R)}{f_{lj}(\rho)}\right|^2~.
\eea

\section{Data systematics} 
\label{sec:syst} 
\setcounter{equation}{0} 
\renewcommand{\theequation}{3.\arabic{equation}} 

At present the amount of neutron emission experimental data from the ground state is very limited, due to the difficulties
to measure half lives of these very unstable nuclei, close to the neutron dripline \cite{Dim21}.
Moreover, the experimental errors are relative large. This is seen from Table I,
where we give the available experimental data concerning quantum numbers, Q-value, half life and error
of neutron emission from ground state. In the last columns are given the corresponding references.
Notice that, except for the first three lines, we assigned angular momenta according to the standard spherical shell model
scheme \cite{Rin80}, due to the fact that this quantity in general is not experimentally determined.
Let us mention here that the standard closure numbers N=2, 8, 20 seems to change with the increase of the asymmetry
$N-Z$. As can be seen from Fig. \ref{fig1} of Ref. \cite{Ume16}, the neutron shell closure at N=20 disapears and therefore
the assignement of the angular momentum may be different from the standard scheme.
Thus, we considered one exception labeled by an asterisk in Table I for $l=1$,
corresponding to a much better fit to the theoretical curve, describing the angular momentum dependence of the half life.

\newpage
\begin{center}
{TABLE I. Neutron experimental emission data: charge, neutron, mass numbers, angular momentum,
Q-value, reduced radius estimated at the nuclear geometrical radius, half life, error and reference.}
\begin{tabular}{|c|c|c|c|c|c|c|c|c|}
\hline
$Z$ & $N$ & $A$ & $l$ & $Q$ (keV)  & $\rho$ &  $T$ (s)    & $\delta T$ (s) & Ref. \cr
\hline
 1  &  3  &  4  &  1  &  1600 & 0.528 & 8.42(-23)  &  - & \cite{Kel92} \cr 
 2  &  3  &  5  &  1  &  735  & 0.386 & 7.89(-22)  &     -      & \cite{Til02} \cr
 2  &  5  &  7  &  1  &  409  & 0.322 & 3.04(-21)  &  0.41(-21) & \cite{Til02}\cr
 4  &  9  & 13  &  2  &  510  & 0.442 & 2.70(-21)  &  1.80(-21) & \cite{Pen95} \cr
 4  & 11  & 15  &  2  & 1800  & 0.870 & 7.87(-22)  &  2.71(-22) & \cite{Sny13} \cr
 5  & 11  & 16  &  1$^*$  &   82  & 0.190 & $>$4.56(-21) &     -    & \cite{Lec09} \cr
 5  & 15  & 20  &  2  & 1560  & 0.892 & 9.12(-22)  &      -     & \cite{Leb18}\cr
 8  & 17 &  25  &  2  &  725  & 0.655 & 2.28(-22)  & $^{+6.84(-22)}_{-2.28(-22)}$ & \cite{Cae13} \cr
 9  & 19  & 28  &  2  &  220  & 0.375 & 4.56(-20)  &      -     & \cite{Chr12}\cr
    &     &     &  1  &  219  & 0.373 & 2.53(-21)  & $^{+0.72(-21)}_{-0.46(-21)}$ & \cite{Rev20}\cr
\hline
\end{tabular}
\end{center}

In the last two columns concerning $^{28}$F we mentioned two different experimental data
with different angular momenta. The value $l=2$ corresponds to the standard spherical shell model
scheme \cite{Rin80}, while $l=1$ evidences "the island of inversion around N=20 which includes $^{28}$F,
and most probably $^{29}$F, and suggests that $^{28}$O is not doubly magic" \cite{Rev20}.
Notice that the half lives for transitions where the experimental errors are not mentioned were estimated
by using the phenomenological R-matrix analysis \cite{Lan58}.

\begin{figure}[h]
\begin{center} 
\includegraphics[width=10cm]{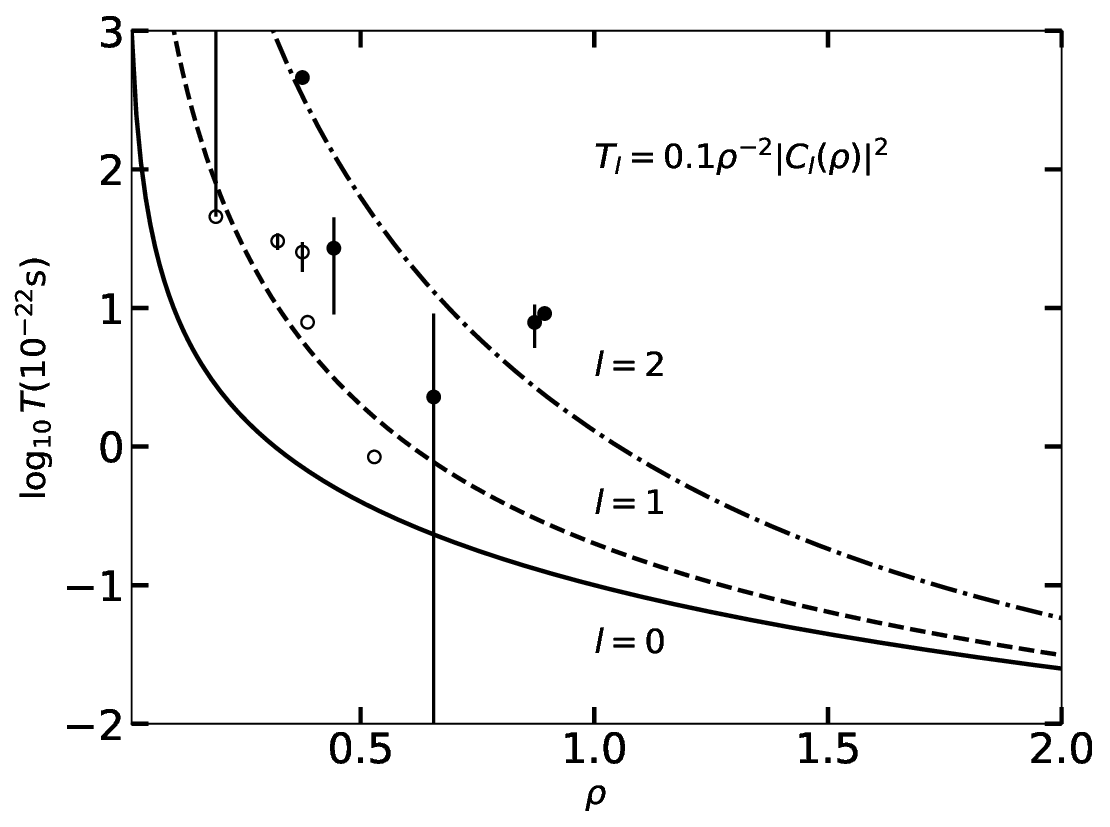} 
\caption{
Experimental half lives in logarithmic scale  (in $10^{-22}$ units) versus reduced radius $\rho$
for $l$=1 (open circles) and $l=2$ (dark circles).
The fitting curves are given by $T_l=0.1~|h^{(+)}_l(\rho)|^2=0.1~|C_l(\rho)|^2/\rho^2$ for angular momenta l=1,2.
We also added the curve corresponding to $l=0$ and
plotted predicted half lives corresponding to the interval $\rho\in [1,2]$.
}
\label{fig2}
\end{center} 
\end{figure}

These nuclei belong to the neutron dripline. They are very unstable, the largest half life
being only by two orders of magnitude above the characteristic nuclear time $10^{-22}$ s.
Let us stress on the fact that the amount of experimental data is very rich if one considers
neutron emission from excited states populated by beta decay, defining the so-called beta delayed neutron emission
\cite{Lia20}. The analysis of these data will be performed in a forthcoming paper.
  
In Fig. \ref{fig2} we plotted the experimental half lives in logarithmic scale (in $10^{-22}$ units) together with the simplest theoretical
prediction given by (\ref{T}), fitting these data by using one common parameter for all angular momenta 
\bea
\label{Tl}
T_l=~a|h^{(+)}_l(\rho)|^2=~a\frac{|C_l(\rho)|^2}{\rho^2}~,
\eea
where the fit parameter $a=0.1$ and the rms error $\sigma = 0.5$.
Thus, we neglected the influence of the factor multiplying $|h^{(+)}_l(\rho)|^2$ in Eq. (\ref{T}).
We notice a reasonable agreement between this prediction and experimental data within the displayed experimental errors.
We also plotted predicted half lives corresponding to the interval $\rho\in [1,2]$.

\begin{figure}[h]
\begin{center} 
\includegraphics[width=10cm]{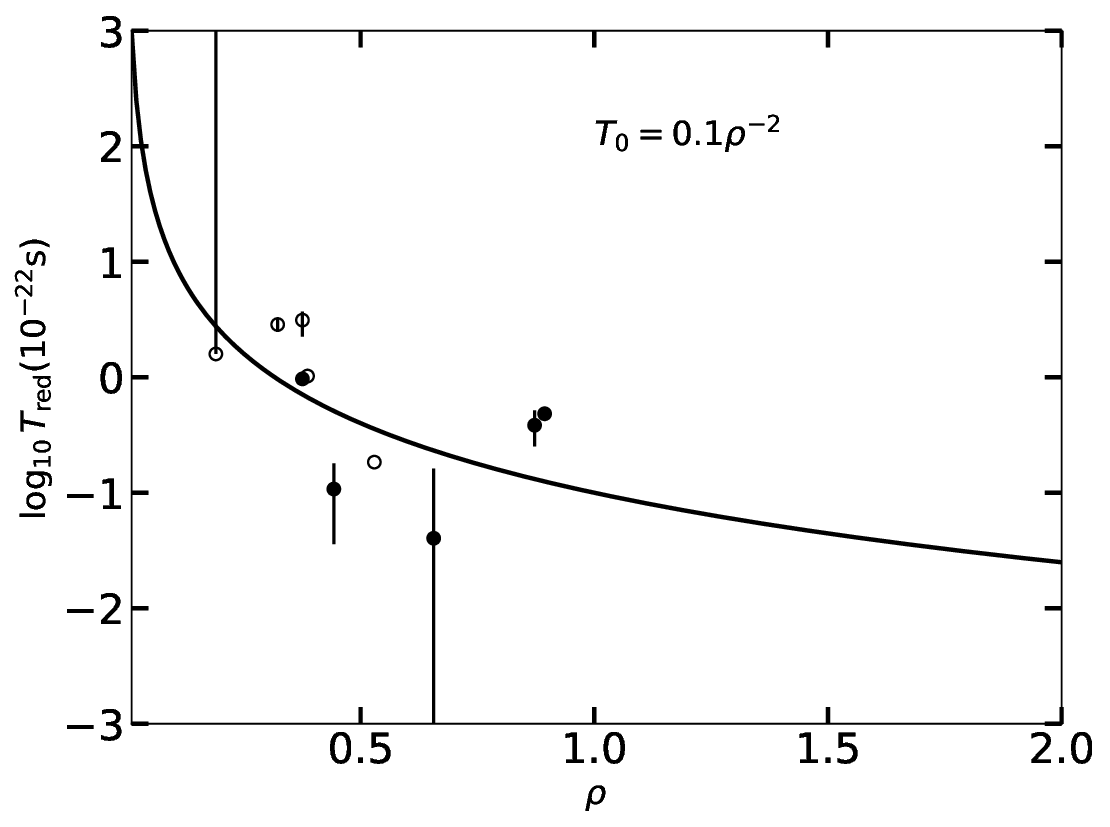} 
\caption{
Experimental half lives in logarithmic scale (in $10^{-22}$ units) versus reduced radius $\rho$
divided by the centrifugal factor $|C_{l}(\rho)|^2$.
The fitting curve represents the same ansatz as in Fig. \ref{fig2} for the monopole case 
$T_0=0.1~|h^{(+)}_0(\rho)|^2=0.1/\rho^2$. 
We plotted predicted half-lives corresponding to the interval $\rho\in [1,2]$.
}
\label{fig3}
\end{center} 
\end{figure}

Let us mention that in Ref. \cite{Del06} the influence of the centrifugal barrier for proton emitters was excluded
by dividing the experimental half life to the centrifugal factor $|C_l(\rho)|^2$. 
Here we will use the same centrifugal factor by defining the reduced half life
\bea
\label{Tred}
T_{red}=\frac{T_{exp}}{|C_{l}(\rho)|^2}~.
\eea
It actually correspons to a dependence of the reduced half life upon the monopole spherical Hankel function squared. 
In Fig. \ref{fig3} we plotted the dependence of the reduced half life in logarithmic scale  (in $10^{-22}$ units)
versus the reduced radius $\rho$, together with the fitting curve 
\bea
\label {T0}
T_0=0.1~|h^{(+)}_0(\rho)|^2=\frac{0.1}{\rho^2}~.
\eea
We obtained a reasonable correlation between this curve and experimental points. 

\begin{figure}[h]
\begin{center} 
\includegraphics[width=10cm]{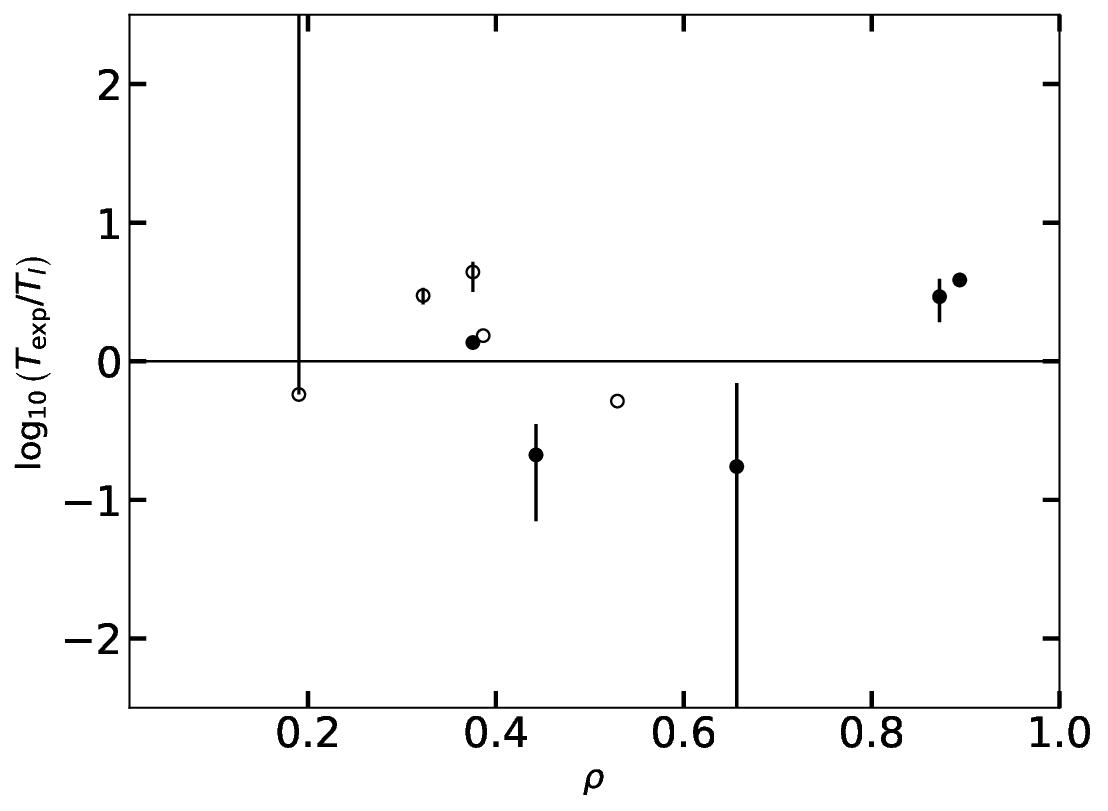} 
\caption{Experimental formation probability given by the ratio $T_{exp}/T_l$, where $T_l$
is gicen by (\ref{Tl}), in logarithmic scale.
}
\label{fig4}
\end{center} 
\end{figure}

Let us here mention that the role of the particle emission formation probability, prportional
to the wave function squared $|f_{lj}(R)|^2$, was extensivelly investigated.
In the recent review \cite{Now21} various versions of the shell model calculations in this region were analyzed, 
starting with the Kuo-Brown code \cite{Kuo66}.
The particle emission processes were analyzed by using the resonant states in a deformed Woods-Saxon 
plus Coulomb mean field \cite{Lio96,Fer97,Qi12}, as well as more sophisticated open quantum system formalism 
for the description of weakly bound nuclei far from the valley of stability \cite{Mic02,Oko03}. 

In order to evidence the role of the neutron wave function we plotted in Fig. \ref{fig4} the so-called
"experimental formation probability", given by the ratio between the experimental half life
and the best fit of the centrifugal penetration factor 
$f_{exp}(\rho)=T_{exp}/[0.1/\rho^2]$.
This quantity is obviously proportional to the neutron wave function squared $|f_{lj}(R)|^2$ entering
the definition of the half life ((\ref{T}).
The result in Fig. \ref{fig4} is nothing else that the ratio between experimental points and the corresponding 
values of the curve in Fig. \ref{fig3}.
Notice that the rms error is $\sigma=0.5$, corresponding to a factor about three in the half life.
Therefore the role of the neutron wave function is not so important for a rough estimate of the neutron 
emission half life for transitions connecting ground states.

\begin{figure}[h]
\begin{center} 
\includegraphics[width=10cm]{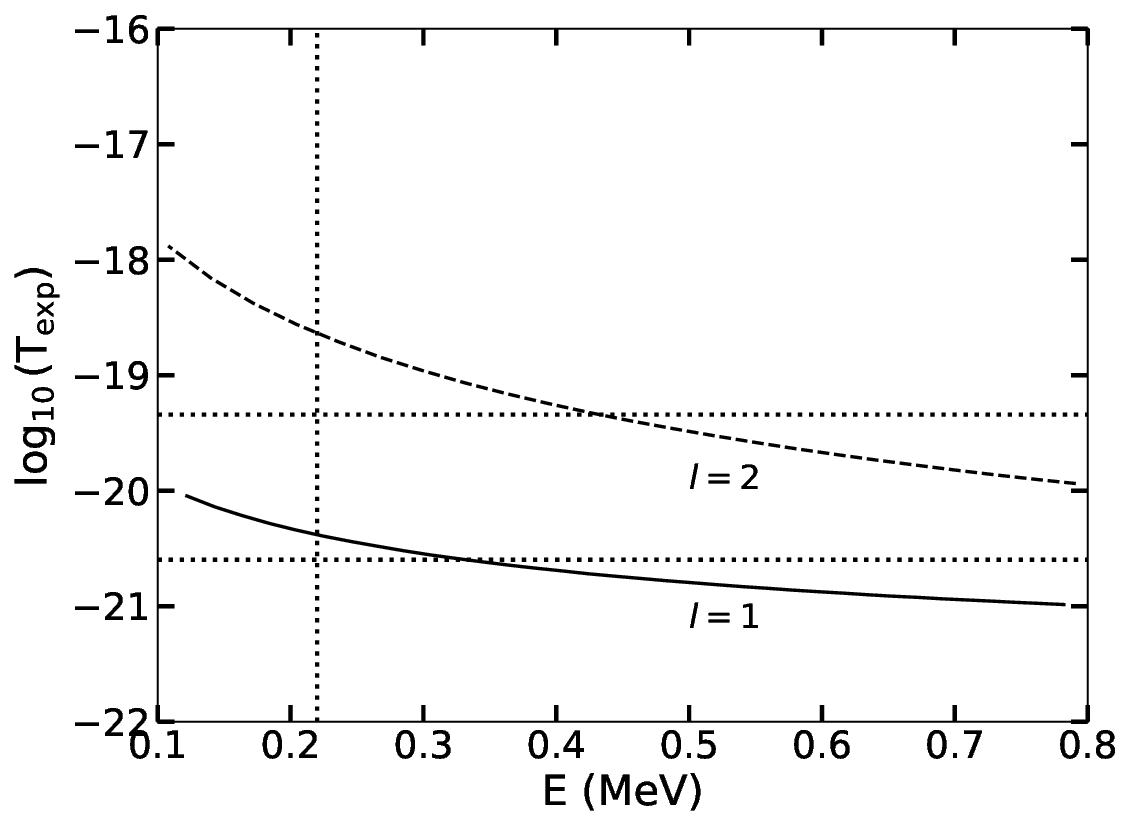} 
\caption{Logarithm of the half life versus energy of l=2 (upper curve)
and l=1 (lower curve) resonant levels in a spherical Woods-Saxion mean field for $^{28}$F. 
The vertical line represents the experimental Q-value and the two horizontal lines 
the experimental half lives given by the last two lines of Table I. 
}
\label{fig5}
\end{center} 
\end{figure}

On the other hand, it turns out that the most sensitive ingredient is the angular momentum of the emitted neutron,
similar to the proton emission systematics \cite{Del06}. This is shown in Fig. \ref{fig2}
and also in Fig. \ref{fig5}, where we plotted the theoretical half life of a resonant state  
above the Fermi surface with angular momentum l=1 (lower curve) and l=2 (lower curve) versus its energy for $^{28}$F.
Notice here two orders of magnitude difference in half lives between these curves at the experimental Q-value. 
These states were generated by using a spherical Woods-Saxon meand field \cite{Del10} 
with universal parametrisation \cite{Dud82,Cwi87} for $^{28}$F.
We used the theoretical decay width given by (\ref{Gamma}) and changed the energy of the resonant
state by using the real part of the central potential. It turns out that this relation
is not sensitive to the change of other parameters of the used nuclear mean field. 
This feature is also general for proton emission processes connecting ground states \cite{Del06a}.
By vertical dotted line we denoted the experimental Q-value and by horizontal dotted lines the two 
experimental values of the half life mentioned in the last two lines of Table I. 
Let us mention that our theoretical prediction  for l=1 is much closer to the experintal value, 
thus confirming the above statement aboout the island of inversion in this region and magicity of $^{28}$O.

\section{Conclusions} 
\label{sec:concl} 
\setcounter{equation}{0} 
\renewcommand{\theequation}{4.\arabic{equation}} 

Concluding, we analyzed the available experimental data concerning the neutron emission half life
between ground states for light nuclei with $A\leq $ 28. 
In the case of charged particles it is well known the correlation between the logarithm of the half life and Coulomb parameter
$\chi\sim Z Q^{-1/2}$, characterizing the outgoing Coulomb-Hankel wave. 
For neutral particles this function is replaced by the spherical Hankel wave.
It turns out that the half lives of light neutron emitters are rather well fitted by the modulus squared 
of the spherical  Hankel wave, estimated at the reduced nuclear radius
$\rho\sim A^{1/3}Q^{1/2}$ corresponding to the experimental Q-value. 
Moreover, as in the case of the proton emission, the reduced half lives estimated by excluding the centrifugal factor 
are proportional to the modulus squared of the monopole spherical Hankel function.
Therefore the well known dependence in the emission of charged particles $\log_{10} T\sim ZQ^{-1/2}$
(Geiger-Nuttall law) is replaced in the emission of neutral particles by $T\sim\rho^{-2}\sim A^{-2/3}Q^{-1}$
for monopole transitions.
This is a remarkable feature taking into accound the relative short half lives of these neutron rich nuclei.
It turns out that the estimated influence of the neutron formation probability is about a factor of three,
while a difference by one unit in the angular momentum leads to an effect about two orders of magnitude
in the half life. Therefore the investigation of the neutron emission can be a powerfull tool to
assign the angular momentum of the unbound neutron.

\ack
This work was supported by the grant of the Romanian Ministry of Education and Research PN 23210101/2023.


\end{document}